\documentclass[conference]{IEEEtran}
\IEEEoverridecommandlockouts
\usepackage{graphicx}  
\usepackage{amsmath, amsopn, psfrag, amsthm}   
\usepackage{amssymb}
\usepackage{color}
\usepackage{algorithm}
\usepackage{algorithmic}
\usepackage[utf8]{inputenc}
\usepackage[english]{babel}

\usepackage{psfrag}
\usepackage{epsfig}
\usepackage{cite}   

\usepackage[letterpaper, left=1.1in, right=1.1in, bottom=1in, top=0.75in]{geometry}

\usepackage{amssymb,amsmath,mathtools}

\usepackage{graphicx}  
\usepackage{amsmath, amsopn}   
\usepackage{amssymb}
\usepackage{color}
\usepackage{psfrag}
\usepackage{epsfig}
\usepackage{stfloats}
\usepackage{lipsum}
\usepackage{multicol}  

\usepackage[overload]{empheq}

\def\bup{{\boldsymbol{\Upsilon}}}

\def\bftheta{{\boldsymbol{\theta}}}
\def\bfchi{{\boldsymbol{\chi}}}
\def\bfmu{{\boldsymbol{\mu}}}
\def\bftheta{{\boldsymbol{\theta}}}

\graphicspath{{./img/}}
\DeclareGraphicsExtensions{.pdf,.jpeg,.png, .eps}
\usepackage{amssymb,amsmath,mathtools}

\begin{document}
\vspace{-0.12in}
\setlength{\columnsep}{.2in}

\title{A Study of Dynamic Multipath Clusters at 60 GHz in a Large Indoor Environment}
\vspace{-0.22in}
\linespread{.94}

\author{
\IEEEauthorblockN{Manijeh Bashar\IEEEauthorrefmark{1}, Katsuyuki Haneda\IEEEauthorrefmark{2}, Alister G. Burr\IEEEauthorrefmark{1}, and Kanapathippillai Cumanan\IEEEauthorrefmark{1}}
 \IEEEauthorblockA{\IEEEauthorrefmark{1}Department of Electronic Engineering, University of York, Heslington, York, UK, 
\\\IEEEauthorblockA{\IEEEauthorrefmark{2}Department of Electronics and Nanoengineering, Aalto
University, Espoo, Finland,}
Email:{ \{mb1465, alister.burr, kanapathippillai.cumanan\}@york.ac.uk}, katsuyuki.haneda@aalto.fi}
}


\vspace{-0.1in}
\linespread{.94}
\maketitle
\vspace{-0.1in}
\linespread{.94}
\begin{abstract}
The available geometry-based stochastic channel models (GSCMs) at millimetre-wave (mmWave) frequencies do not necessarily retain spatial consistency for simulated channels, which is essential for small cells with ultra-dense users. In this paper, we work on cluster parameterization for the COST 2100 channel model using mobile channel simulations at 61 GHz in Helsinki Airport. The paper considers a ray-tracer which has been optimized to match measurements, to obtain double-directional channels at mmWave frequencies. A joint clustering-tracking framework is used to determine cluster parameters for the COST 2100 channel model. The KPowerMeans algorithm and the Kalman filter are exploited to identify the cluster positions and to predict and track cluster positions respectively. The results confirm that the joint clustering-and-tracking is a suitable tool for cluster identification and tracking for our ray-tracer results. The movement of cluster centroids, cluster lifetime and number of clusters per snapshot are investigated for this set of ray-tracer results. Simulation results show that the multipath components (MPCs) are grouped into clusters at mmWave frequencies.
\\
$~~${$~~${\textbf{$~~$\textit{$~\textcolor{white}{cc}~$Index terms}}}— Cluster identification, Kalman filter, KPowerMeans, millimetre wave, multi path components.}
\end{abstract}
\section{Introduction}
 \let\thefootnote\relax\footnotetext{The work of A. G. Burr and K. Cumanan was supported by H2020- MSCA-RISE-2015 under grant number 690750. The work on which this paper is based was carried out in collaboration with COST Action CA15104 (IRACON).}
Over the past few years, an abundance of techniques have been proposed as a means to efficiently scale the wireless capacity. It remains unclear which technology or set of technologies can meet the demand. One promising set of technologies for the 5th Generation (5G) cellular network is reviewed in \cite{multidebbah}: the combination of large antenna arrays and short wavelength carrier waves. This combination allows for a greater bandwidth availability and extremely high spectral efficiency by utilizing a large number of antennas, whilst occupying a relatively small area. This technology is known as Massive multiple-input multiple-output (MIMO) in the millimeter-wavelength (mmWave) spectrum \cite{CiareJointspatial}.

Most standardized MIMO channel models such as IEEE 802.11 \cite{Molish_tufvesson} and the most recent 3GPP channel model \cite{3GPP_spec} rely on clustering \cite{Molish_tufvesson}. The same applies to the recent COST channel models, e.g., the COST 2100 model \cite{katsumimocost,ouriet_mic,our_ew,ourtvt18}. These models are geometry-based stochastic channel models (GSCMs) that are mathematically tractable, though to a limited extent, to investigate the performance of MIMO systems \cite{Rappaport_globe15}. The concept of clustering is an essential basis of GSCMs to characterize scatterers in the cell environments. In \cite{katsu_vtc15_winner,katsu_icc16,katsu_ref_Iqbal,katsu_ref_Schneider,Gustafson_katsu}, the authors use clusters to characterize measured multipath channels for a GSCM in mmWave bands. The available GSCMs at mm-waves do not necessarily retain the spatial consistency of simulated channels due to lack of cluster dynamics, which is essential for small cells with ultra-dense users.  In this paper, we work on cluster parameterization to investigate the spatial consistency, using a ray-tracer which is adjusted to produce results consistent with measurements.

Unlike previously available clustering algorithms, in this paper the coordinates are exploited for which the multipath components (MPCs) interact with surrounding objects for a fixed position of mobile station (MS) and base station (BS). To the best of our knowledge, previously clustering has been performed in a double-directional setting, i.e., considering both angle of arrival (AoA) and angle of departure (AoD). A consistent scheme to identify and track clusters based on the spatial coordinates of the MPCs (the $[x,y,z]$-coordinates of the MPCs) is presented. To investigate the performance of the proposed clustering scheme we exploit a set of ray-tracer results in Helsinki's airport described in \cite{katsu_airpot_vtc18}, which is very accurate to present the propagation properties such as specular reflections, diffraction, diffuse scattering \cite{katsu_trans_16_laser}.
The contributions of the paper are summarized as follows: 
\begin{itemize}
\item[\textbf{1.}] We study whether clusters exist or not.
\item[\textbf{2.}] For the first time, we perform clustering of dynamic multipath channels.
\item[\textbf{3.}] $[x,y,z]$ coordinate-based clustering.
\end{itemize}
\subsection{Outline}
The rest of the paper is organized as follows. Section II describes
the ray-tracer and simulation area, and Section III provides the MPC clustering-and-tracking framework. The simulation results and discussion are presented in Section IV while Section V concludes the paper.
\subsection{Notation}
The following notations are adopted in the rest of the paper. Uppercase and lowercase
boldface letters are used for matrices and vectors, respectively.
The notation $|x|$, $|\textbf{X}|_{\text{det}}$ and $|\textbf{x}|_{\text{size}}$ stand for the absolute value of $x$, determinant of matrix $\textbf{X}$, and the size of vector $\textbf{x}$, respectively. $\textbf{X}^{-1}$ and $\textbf{X}^{T}$ 
denote the inverse and transpose of matrix $\textbf{X}$, respectively. Moreover, $\textbf{I}_n$ introduces identity matrix with size $n\times n$. The Kronecker product of $\textbf{X}$ and $\textbf{Y}$ is presented by $\textbf{X} \otimes\textbf{Y}$.
\section{The Ray-tracer and Simulation Area}\label{raytracer}
\begin{figure}[t!]
\center
\includegraphics[width=79mm]{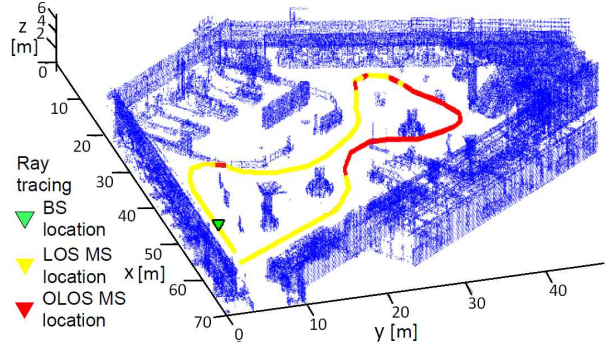}
\caption{Floor plan of the small-cell site in Helsinki airport. For this simulation set-up $f_c=61$ GHz, $BW=2$ GHz refer to the carrier frequency and bandwidth, respectively. Moreover, the position of BS is fixed (the green triangle), while we investigates 2639 positions for MS (the yellow and red points demonstrate the LOS and OLOS, respectively. The total MS route is 132 m, and channels simulated at every 5 cm.}
\label{airport}
\end{figure}

The in-house ray-tracer simulates multipath channels for a large number of links between BS and MS \cite{katsu_airpot_vtc18}. Note that our ray-tracer
works with accurate descriptions of the environment in the form
of point clouds, obtained by laser scanning, and has the ability of simulating relevant propagation properties such as specular reflections, diffraction, diffuse scattering and shadowing \cite{katsu_trans_16_laser}. For more details on our ray-tracer refer to \cite{katsu_airpot_vtc18, katsu_trans_16_laser}.
A check-in hall of Helsinki airport as a representative small-cell scenario is considered as shown in Fig. \ref{airport}. Exploiting the ray-tracer parameters in Fig. \ref{airport}, we obtain the MPCs for links defined by BS and MS locations as in Fig. \ref{airport}. The BS is located 1 m from a wall at a height of 5.7 m whereas the MS is placed at a height of 1.5 m at every 5 cm over a route. In total, 2639 links including 1816 line-of-sight (LOS) and 823 obstructed LOS (OLOS) are simulated. As the ray-tracer calculates interactions of MPC with physical objects in the environments, we save the first and last MPC interacting coordinates $[x, y, z]$ instead of the angle of departure and arrival of each MPC. We assume downlink where BS transmits and MS receives radio signals. The first and last interacting coordinates are the same for a single-bounce path, and are different for a multiple-bounce path. The ray-tracer also derives a complex gain for each MPC.

\section{Clustering-and-Tracking Framework}
Similar to standard clustering algorithms \cite{nicki_track_china06,nicki_indoor_china06}, we independently perform clustering at each snapshot and thereafter the clusters are tracked. Consider $n=1,\cdots,N$ data windows, where at each data window we have $L^{(n)}$ MPCs. Next, we define for each MPC $\textbf{v}_{1,l}^{(n)}\!=\![x_{MS,l}^{(n)},y_{MS,l}^{(n)},z_{MS,l}^{(n)}]$ (the position of MPCs from MS side) and $\textbf{v}_{2,l}^{(n)}=[x_{BS,l}^{(n)},y_{BS,l}^{(n)},z_{BS,l}^{(n)}]$ (the position of MPCs from BS side), and finally we have
\begin{IEEEeqnarray}{rCl}\chi_{l}^{(n)}&=&\left[\textbf{v}_{1,l}^{(n)}\right]=\left[x_{MS,l}^{(n)},y_{MS,l}^{(n)},z_{MS,l}^{(n)}\right].\!\!\!
\end{IEEEeqnarray}
The same equality hold for the BS-side components. This enables us after visualising clusters to plot clusters separately for $v_{1,l}^{(n)}$ and $v_{2,l}^{(n)}$ in physical three-dimensional space as well as defining the matrix $\bfchi^{(n)}=[\chi_{1}^{(n)},\cdots,\chi_{L}^{(n)}]$ Moreover, the $l$th MPC in window $n$ has a power represented by $p_l^{(n)}$ which enables us to define the power vector $\textbf{p}^{(n)}=[p_1^{(n)},\cdots,p_L^{(n)}]$.  
\subsection{Cluster Parameters}
In next step, we define the following parameters for each cluster:
\begin{itemize}
\item[\textbf{1.}] Cluster ID $c$.
\item[\textbf{2.}] Cluster power at time $n$: $\gamma_{c}^{(n)}=\sum_{l\in I_{c}^{(n)}}p_{l}^{n}$, where $I_{c}^{(n)}$ denotes the set of MPCs belonging to cluster $c$ at time $n$.
\item[\textbf{3.}] Total number of MPCs in cluster $c$ at time $n$: $L_{c}^{(n)}=|I_{c}^{(n)}|_{\text{size}}$.
\item[\textbf{4.}] Cluster centroid position:
\begin{small}
\begin{IEEEeqnarray}{rCl}
&&\bfmu_{c}^{(n)}=\left[x_{MS,c}^{(n)},y_{MS,c}^{(n)},z_{MS,c}^{(n)}\right]^T=\frac{1}{\gamma_{c}^{(n)}} \\
&&\left[\!\!\sum_{l\in I_{c}^{(n)}}p_{l}^{n}x_{MS,l}^{(n)},\sum_{l\in I_{c}^{(n)}}p_{l}^{n}y_{MS,l}^{(n)},\sum_{l\in I_{c}^{(n)}}p_{l}^{n}z_{MS,l}^{(n)}\!\!\right]^T\!\!\!\!.\nonumber
\end{IEEEeqnarray}
\end{small}
\item[\textbf{5.}] Combined cluster centroid position and speed:
\begin{small}
\begin{IEEEeqnarray}{rCl}
&&\bftheta_{c}^{(n)}=\\
&&\left[\!\!x_{MS,c}^{(n)},\Delta x_{MS,c}^{(n)},y_{MS,c}^{(n)},\Delta y_{MS,c}^{(n)},z_{MS,c}^{(n)},\Delta z_{MS,c}^{(n)}\!\!\right]^T\!\!\!\!.\nonumber
\end{IEEEeqnarray}
\end{small}
\item[\textbf{6.}] Cluster spread matrix:
\begin{small}
\begin{IEEEeqnarray}{rCl}
\textbf{C}_c^{(n)}=\frac{\sum_{l\in I_{c}^{(n)}} p_l^{(n)}\left(\bfchi_{l}^{n}-\bfmu_{c}^{{n}}\right)\left(\bfchi_{l}^{n}-\bfmu_{c}^{{n}}\right)^T}{\gamma_{c}^{(n)}}\!\!\!.
\end{IEEEeqnarray}
\end{small}
\end{itemize}
Next, similar to terminology in \cite{nicki_track_china06}, a Kalman filter \cite{kay} is used to both track and predict the cluster positions over time. Moreover, an initial-guess process introduces an appropriate initial guess for cluster centroids, and finally the clustering algorithm determines the clusters in the ray-tracer results exploiting the initial guess. 
\subsection{Kalman Filter to Track and Predict Cluster Positions}
We exploit the cluster centroid positions and cluster centroid  speeds for the Kalman tracking \cite{kay}. The following state equations are used:
\begin{subequations}
\label{satet}
    \begin{empheq}
    [left=\empheqlbrace\,]
    {align}
      &
      \bftheta _c^{(n)} = \textbf{A}  \bftheta _c^{(n-1)} + \textbf{B}^{(n)},
      \\&
      \textbf{A} = \textbf{I}_3 \otimes \begin{bmatrix}
    1       & 1 \\
   0      & 1
\end{bmatrix}
      \\&
      \bfmu _c^{(n)}= \textbf{D}  \bftheta _c^{(n)} + \textbf{E}^{(n)},
      \\&
              \textbf{D} = \textbf{I}_3 \otimes \begin{bmatrix}
    1       & 0
\end{bmatrix},
    \end{empheq}
\end{subequations}
where $\textbf{B}^{(n)}$ and $\textbf{E}^{(n)}$ refer to the state-noise with covariance matrix $\textbf{Q}$ and the observation-noise with covariance matrix $\textbf{R}$, respectively. Note that $ \bfmu_c^{(n)}$ introduces the observed cluster centroid position. The prediction and update equations are given by
\begin{subequations}
\label{satet}
    \begin{empheq}
    [left=\text{Prediction}\empheqlbrace\,]
    {align}
      &
      \bftheta _c^{(n|n-1)} = \textbf{A}  \bftheta _c^{(n-1|n-1)},
      \\&
      \textbf{M}^{(n|n-1)} = \textbf{A}  \textbf{M}_c^{(n-1|n-1)}+\textbf{Q},
    \end{empheq}
and update
\end{subequations}
\begin{small}
\begin{subequations}
\label{satet}
    \begin{empheq}
    {align}
      &
      \!\!\textbf{K}^{(n|n)}=\textbf{M}_c^{(n|n-1)}\textbf{D}^T\left(\textbf{D}\textbf{M}^{(n|n-1)}\textbf{D}^T+
      \textbf{R}\right)^{-1}\!,\!\!
      \\&
     \!\!\bftheta_c^{(n|n)}=\bftheta_c^{(n|n-1)}+ \textbf{K}^{(n|n)}\left(\bfmu_c-\textbf{D} \bftheta_c^{(n|n-1)}\right)\!,\!\!
     \\&
     \!\!\textbf{M}^{(n|n)} = \left(\textbf{I}-\textbf{K}^{(n|n)}\textbf{D}\right)\textbf{M}^{(n|n-1)}\!\!
    \end{empheq}
\end{subequations}
\end{small}
\subsection{Association of Clusters}
Association of predicted targets to identified targets is a substantial challenge in any multi-target tracking \cite{nicki_track_china06}. Based on \cite{nicki_track_china06}, the distance between a cluster with parameters $(\bfmu_c, \emph{C}_c)$ and a cluster with centroid $\tilde{\bfmu}$ is called the closeness function and is given by
\begin{IEEEeqnarray}{rCl}
&&d_c\left(\tilde{\bfmu}|\bfmu_c,\textbf{C}_c\right)=\frac{1}{\left(2\pi\right)^\frac{3}{2}\left|\textbf{C}_c\right|_{\text{det}}^\frac{1}{2}}\\&&\exp\left(-\frac{1}{2}\left(\tilde{\bfmu}-\bfmu_c^T\right)^T\textbf{C}_c^{-1}.\left(\tilde{\bfmu}-\bfmu_c^T\right)\right),\nonumber
\end{IEEEeqnarray}
First, the closeness function between the old clusters (with the old covariance matrix) and new centroids and the closeness function between the new clusters (with the old covariance matrix) and old centroids are calculated. Next, for each new cluster the closest old cluster and for each old cluster the closest new cluster is determined. Note that the closest cluster is determined by finding the maximum value of the closeness function. If the closeness function from both directions are exactly the same, these two clusters are associated and assumed to be one cluster. The clusters which are not associated are assumed to be new ones.
\subsection{Initial Guess for Clusters}
The initial guess of the cluster centroids is a challenging task in clustering algorithms. In \cite{nicki_track_china06}, the authors propose a novel initial guess to maximize the distances between the cluster centroids. If there is no cluster prediction available, the path having the strongest power is selected as the first centroid $\hat{\bfmu}_1$ whereas for the case of available cluster prediction, the initial-guess centroid from the prediction is to be as the current initial guess. Note that the multipath component distance (MCD) in this paper is different from  the one used in \cite{nicki_track_china06, nicki_letter_mcd}. The distance measure between MPCs $i$ and $j$ is given by 
\begin{IEEEeqnarray}{rCl}
&&\text{MCD}_{ij}=\\
&&\sqrt{||\text{MCD}_{x_{MS},ij}||^2+||\text{MCD}_{y_{MS},ij}||^+||\text{MCD}_{z_{MS},ij}||^2}.\nonumber
\label{mcd1}
\end{IEEEeqnarray}
Note that in (\ref{mcd1}) we have
\begin{IEEEeqnarray}{rCl}
\label{mcd2}
\text{MCD}_{x_{MS},ij} = \frac{|x_{MS,i}-x_{MS,j}|}{\Delta x_{MS,\text{max}} },
\end{IEEEeqnarray}
where $\Delta x_{MS,\text{max}}=\max{\{|x_{MS,i}-x_{MS,j}|\}}$, and the other terms in (\ref{mcd1}) are evaluated is a the similar way to (\ref{mcd2}).
Next, the weighted distance matrix $\bup \in C^{l\times c}$ between all paths and all initial-guess centroids is evaluated as follows:
\begin{IEEEeqnarray}{rCl}
\label{mcd3}
\bup\left(\bfchi_{l}^{n}-\hat{\bfmu}_{c}\right)=\log_{10}\left(p_{l}^{(n)}\right)\text{MCD}\left(\bfchi_{l}^{n}-\hat{\bfmu}_{c}\right).
\end{IEEEeqnarray}
Following the terminology in \cite{nicki_track_china06}, we select the path with the maximum minimum distance to any centroid as follows:
\begin{IEEEeqnarray}{rCl}
\label{mcd4}
l_{\text{sel}}=\max_l\left\{\min_c\left\{\bup\right\}\right\}.
\end{IEEEeqnarray}
We then assign all MPCs to their closest centroid and cluster power is evaluated. If we do not achieve the maximum number of clusters, and centroid powers are larger than $0.01\%$ of the total snapshot power, we repeat the calculation of the weighted distance matrix $\bup \in \mathbb{C}^{l\times c}$ in (\ref{mcd3}). Otherwise, the last centroid is ignored and the algorithm is stopped.
\subsection{Clustering Algorithm}
The KPowerMeans clustering algorithm is investigated in \cite{nicki_vtc06_clustering_frame}, and it performs as follows: the initial-guess algorithm is applied, and the KPowerMeans clustering algorithm is run only once as the initial guess as are constant. For more details on the KPowerMeans clustering algorithm refer to \cite{nicki_vtc06_clustering_frame}. Note that if any cluster occupies less than $1\%$ of total cluster power, we re-start the clustering algorithm with the initial guess, with the number of clusters is reduced by one. Therefore, it is possible that the algorithm ends with a single cluster. 

\begin{figure}[t!]
\center
\includegraphics[width=83mm]{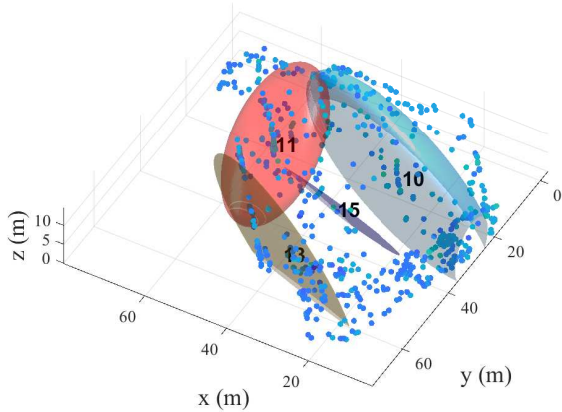}
\vspace{-.72cm}
\caption{Tracked Rx-side clusters in Helsinki airport in snapshot 3.}
\label{cluster_rx_snap1}
\end{figure}
\begin{figure}[t!]
\center
\includegraphics[width=83mm]{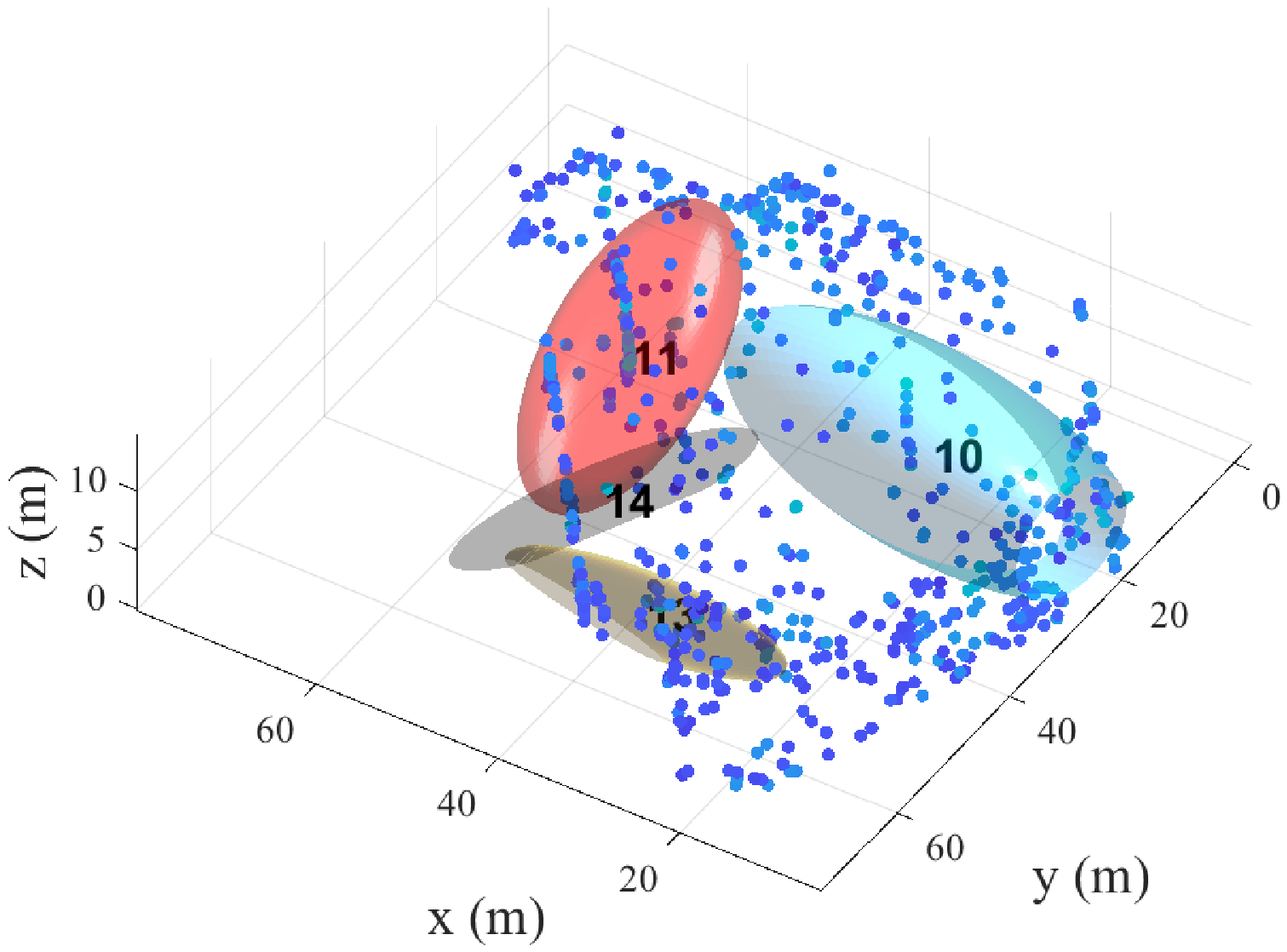}
\vspace{-.1cm}
\caption{Tracked Rx-side clusters in Helsinki airport in snapshot 4.}
\label{cluster_rx_snap2}
\end{figure}
\begin{figure}[t!]
\center
\includegraphics[width=83mm]{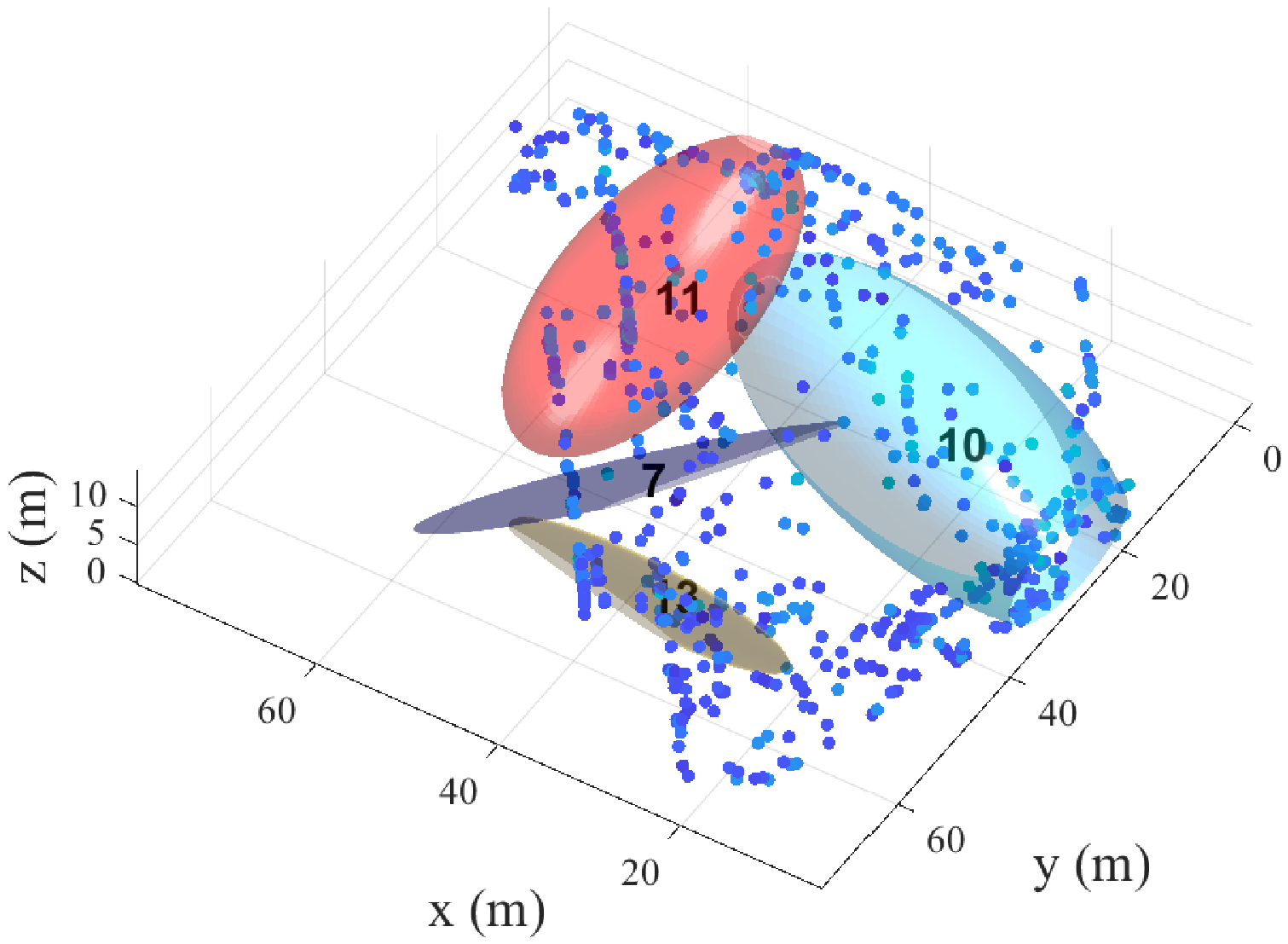}
\vspace{-.1cm}
\caption{Tracked Rx-side clusters in Helsinki airport in snapshot 5.}
\label{cluster_rx_snap3}
\end{figure}
\begin{figure}[t!]
\center
\includegraphics[width=83mm]{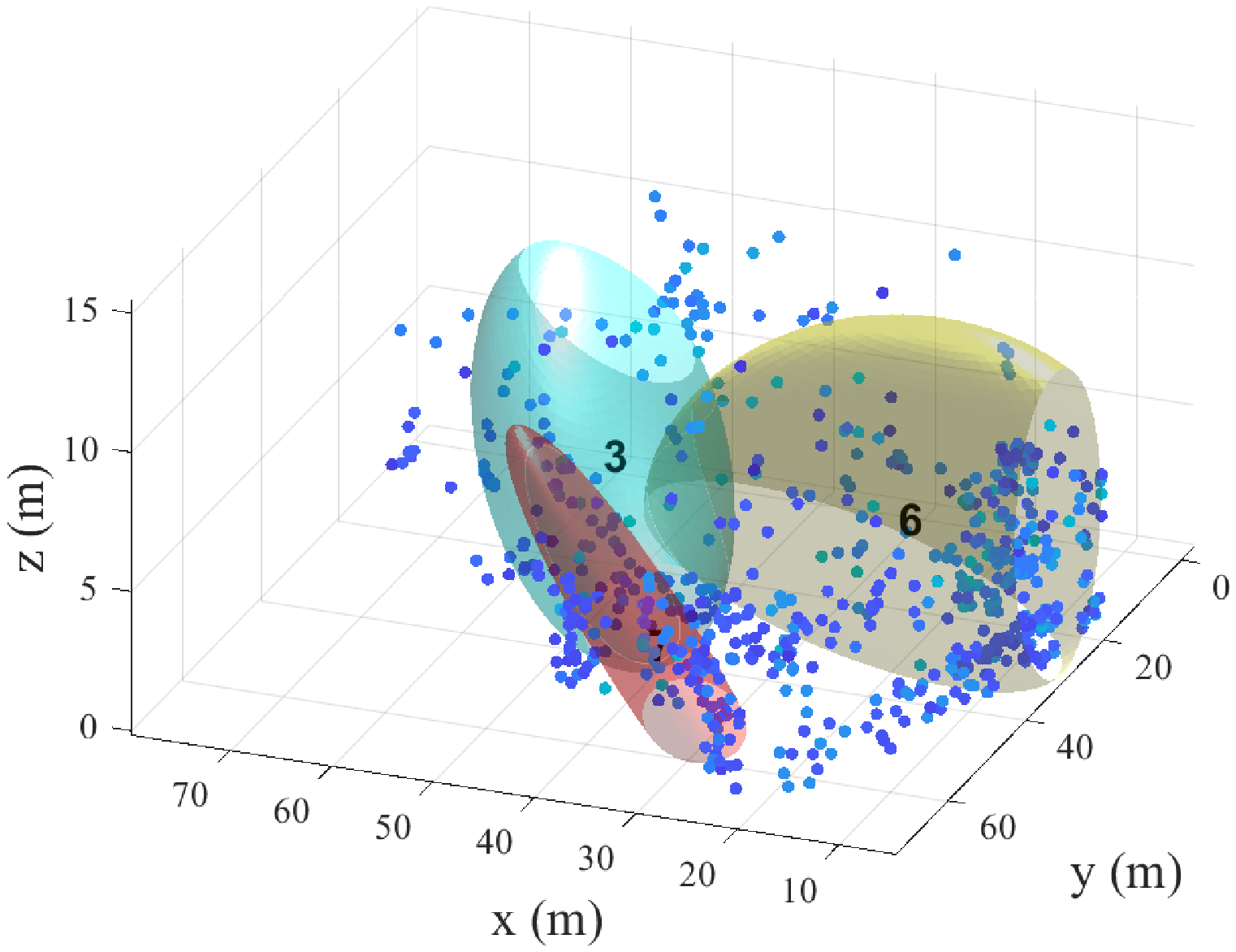}
\vspace{-.1cm}
\caption{Tracked Tx-side clusters in Helsinki airport in snapshot 12.}
\label{cluster_tx_snap1}
\end{figure}
\section{Results and Discussion}
\begin{figure}[t!]
\center
\includegraphics[width=83mm]{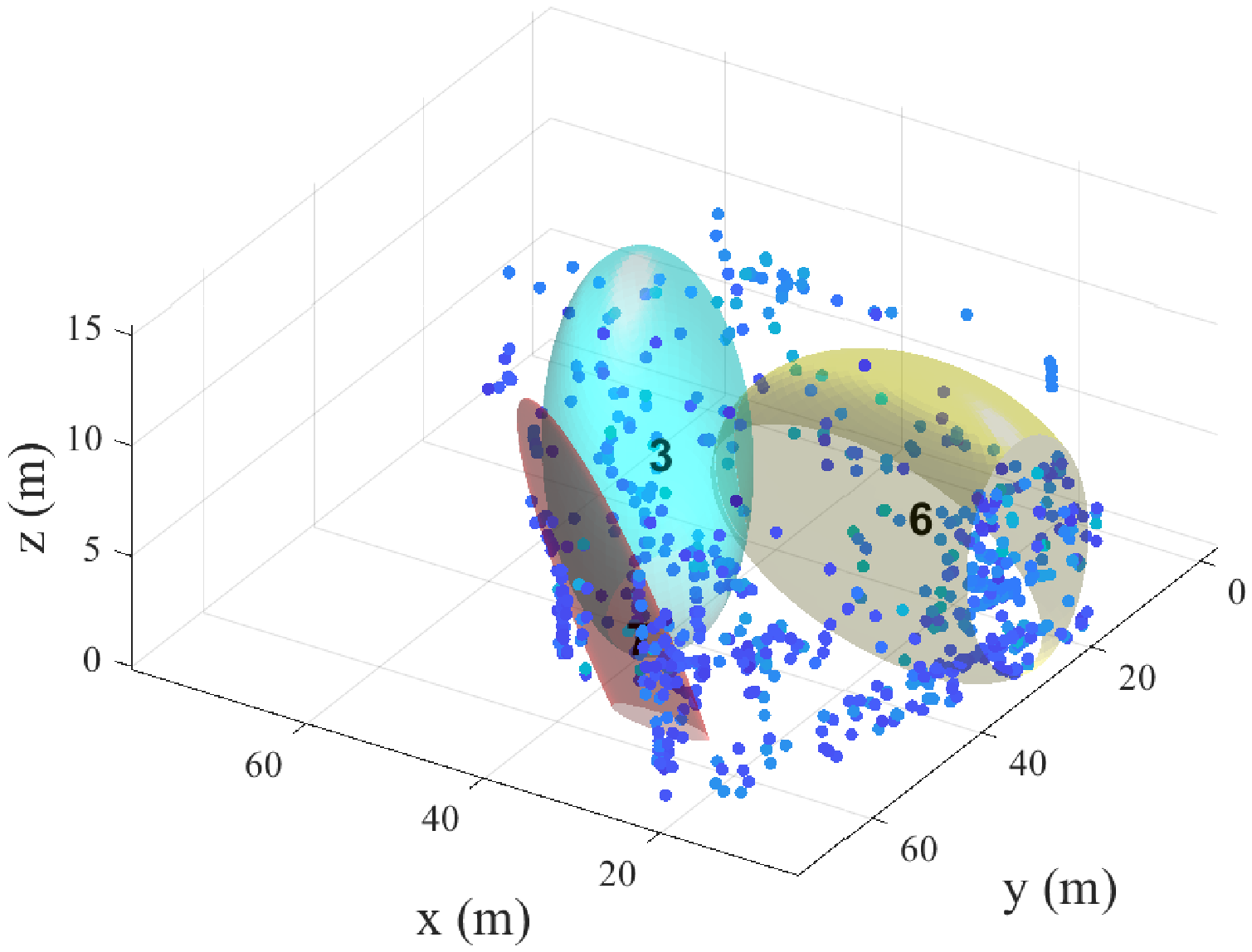}
\vspace{-.1cm}
\caption{Tracked Tx-side clusters in Helsinki airport in snapshot 13.}
\label{cluster_tx_snap2}
\end{figure}

\begin{figure}[t!]
\center
\includegraphics[width=82mm]{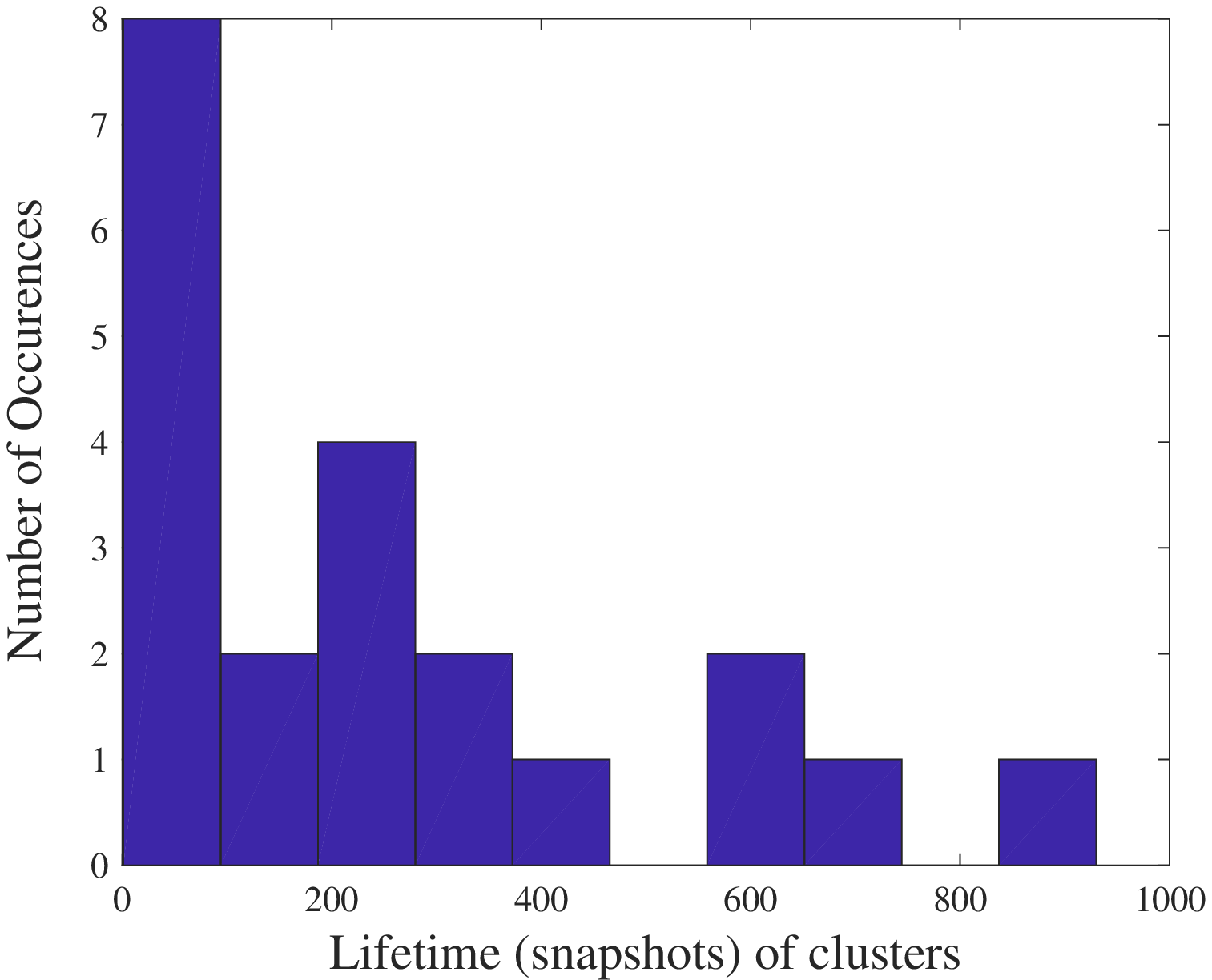}
\vspace{-.1cm}
\caption{Histogram of Rx-side clusters cluster lifetimes (snapshots).}
\label{hist_liftetime_rx}
\end{figure}
\begin{figure}[t!]
\center
\includegraphics[width=82mm]{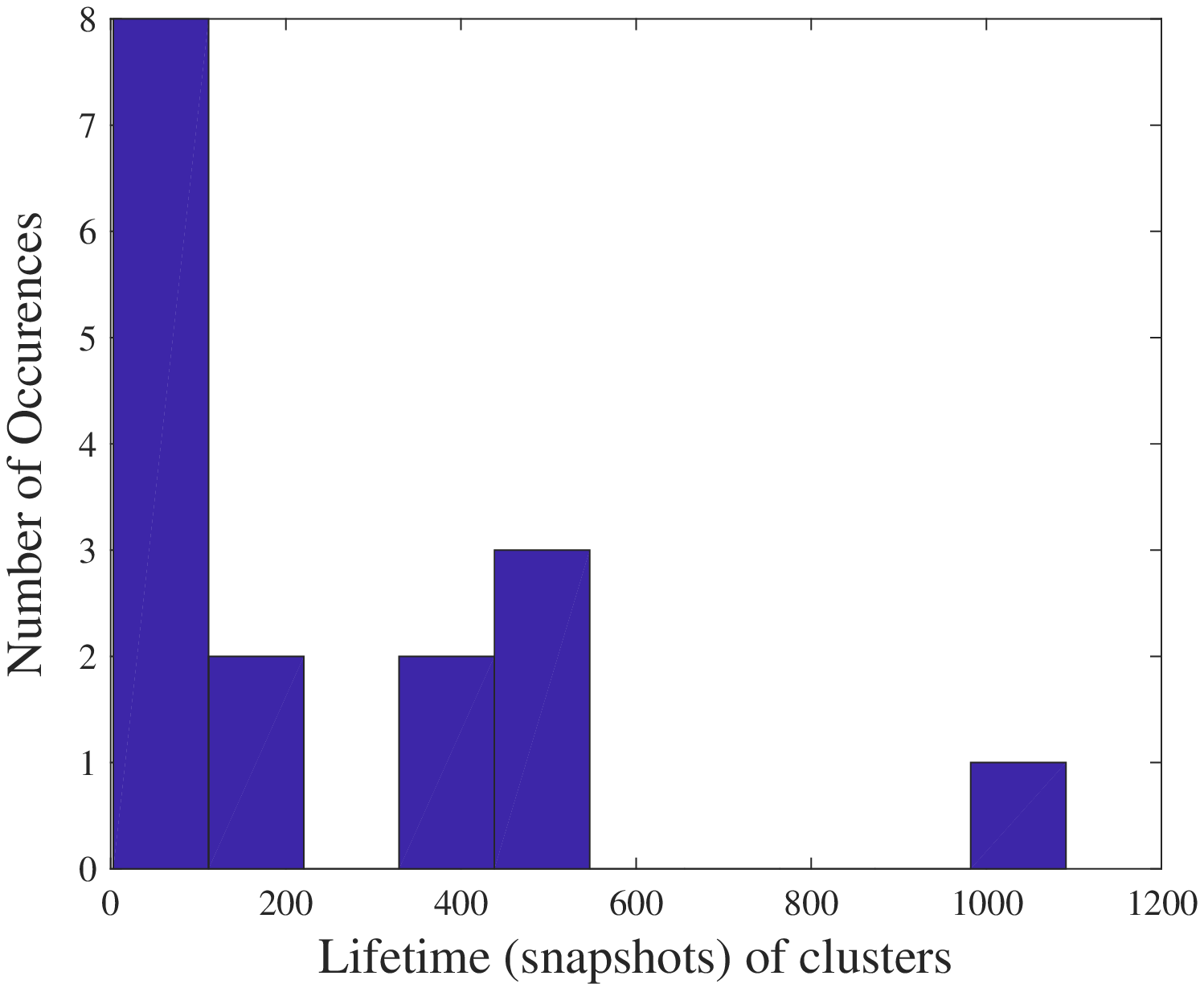}
\vspace{-.2cm}
\caption{Histogram of Tx-side cluster lifetimes (snapshots).}
\label{hist_liftetime_tx}
\end{figure}

\begin{figure}[t!]
\center
\includegraphics[width=82mm]{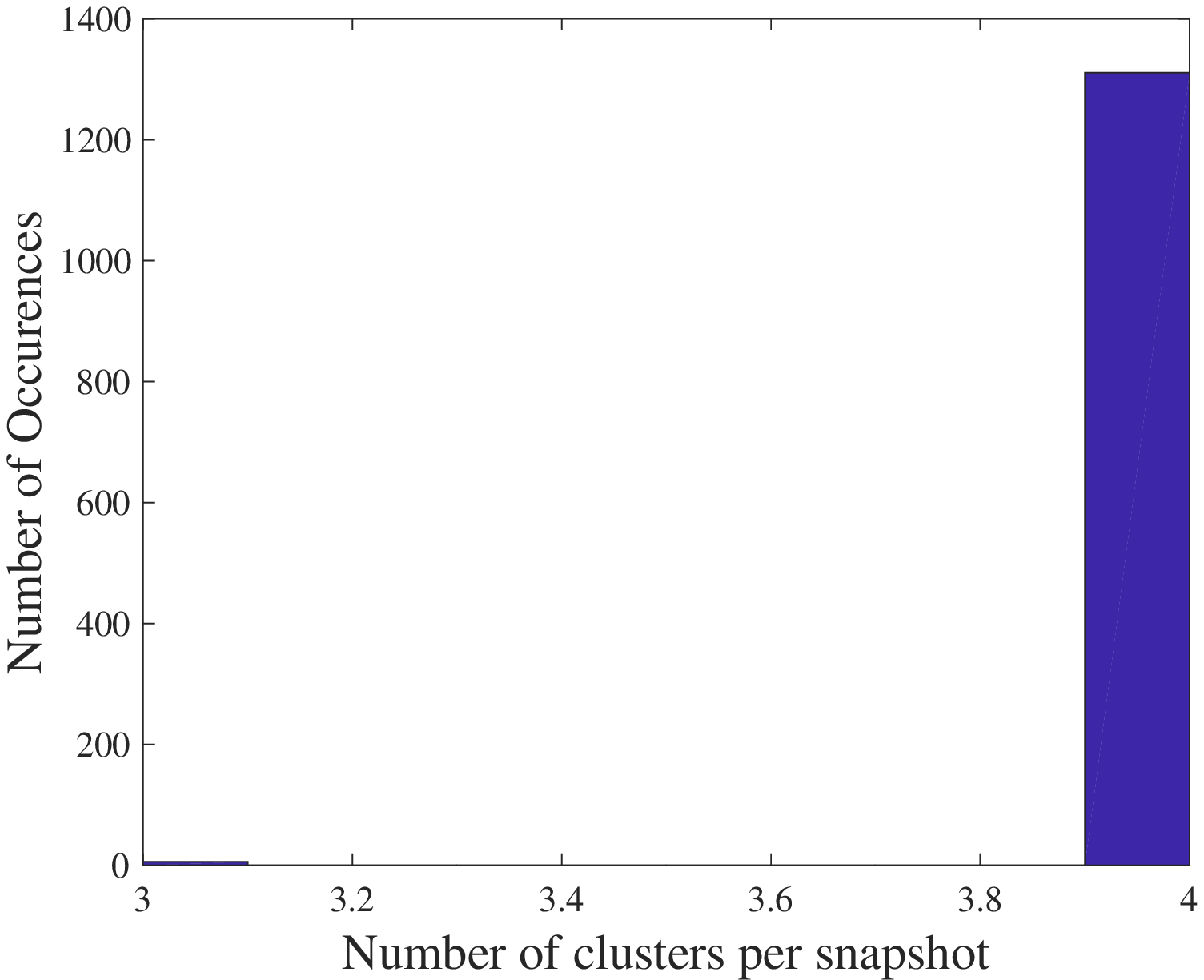}
\vspace{-.2cm}
\caption{Histogram of Rx-side cluster lifetimes (snapshots).}
\label{hist_nc_rx}
\end{figure}
\begin{figure}[t!]
\center
\includegraphics[width=82mm]{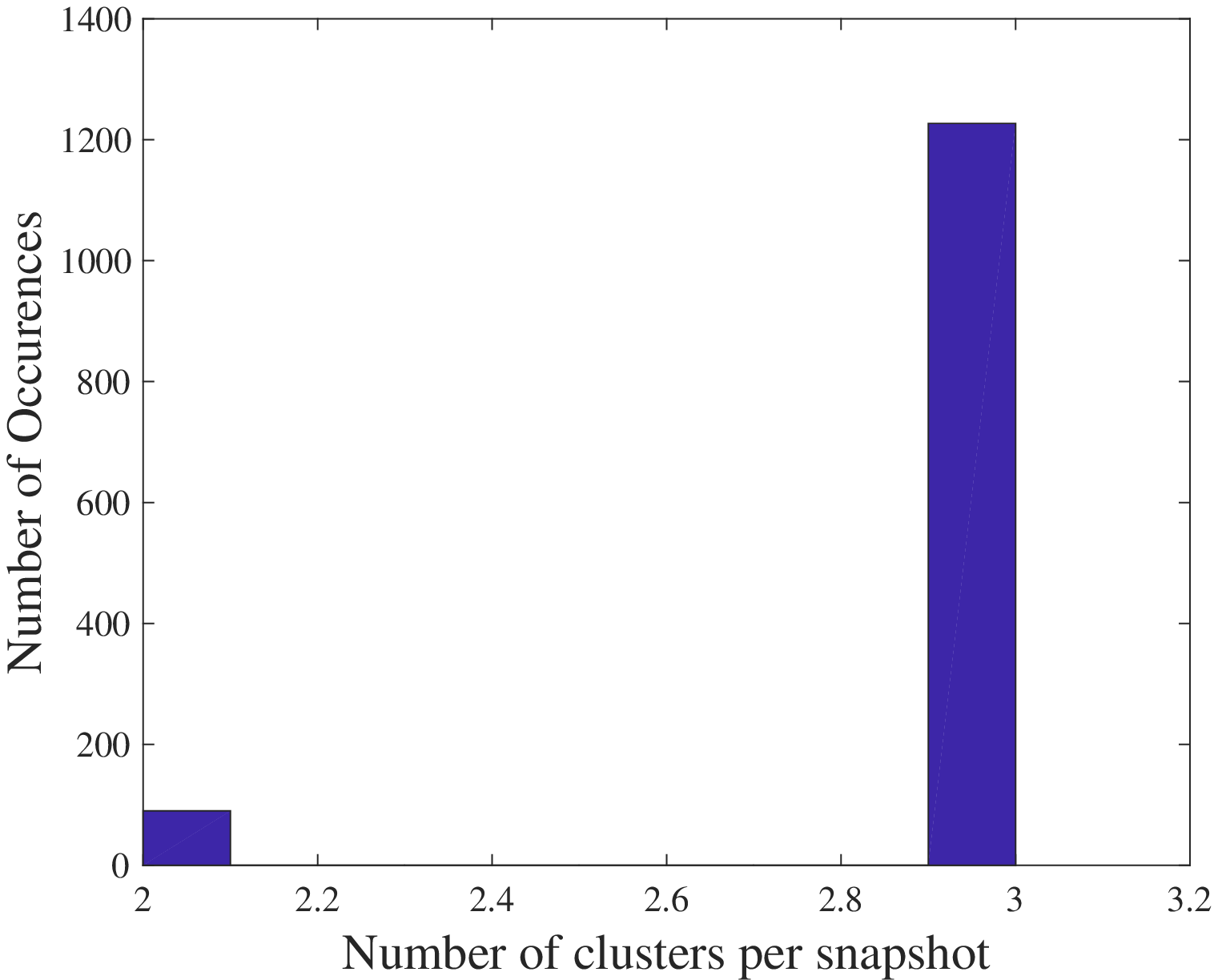}
\vspace{-.2cm}
\caption{Histogram of Tx-side cluster lifetimes (snapshots) .}
\label{hist_nc_tx}
\end{figure}
\begin{figure}[t!]
\center
\includegraphics[width=82mm]{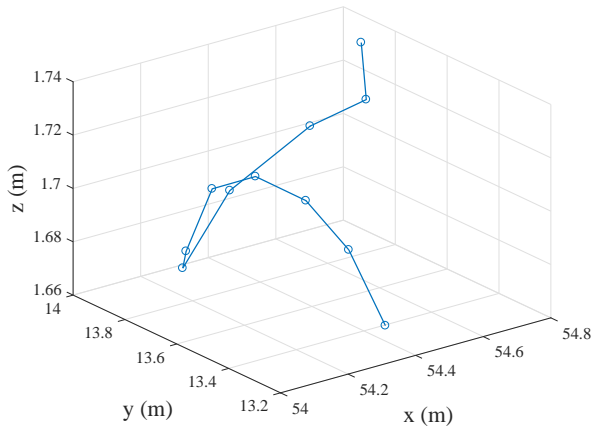}
\vspace{-.2cm}
\caption{Tracked centroid of exemplary moving cluster.}
\label{center_rx}
\end{figure}

\begin{figure}[t!]
\center
\includegraphics[width=82mm]{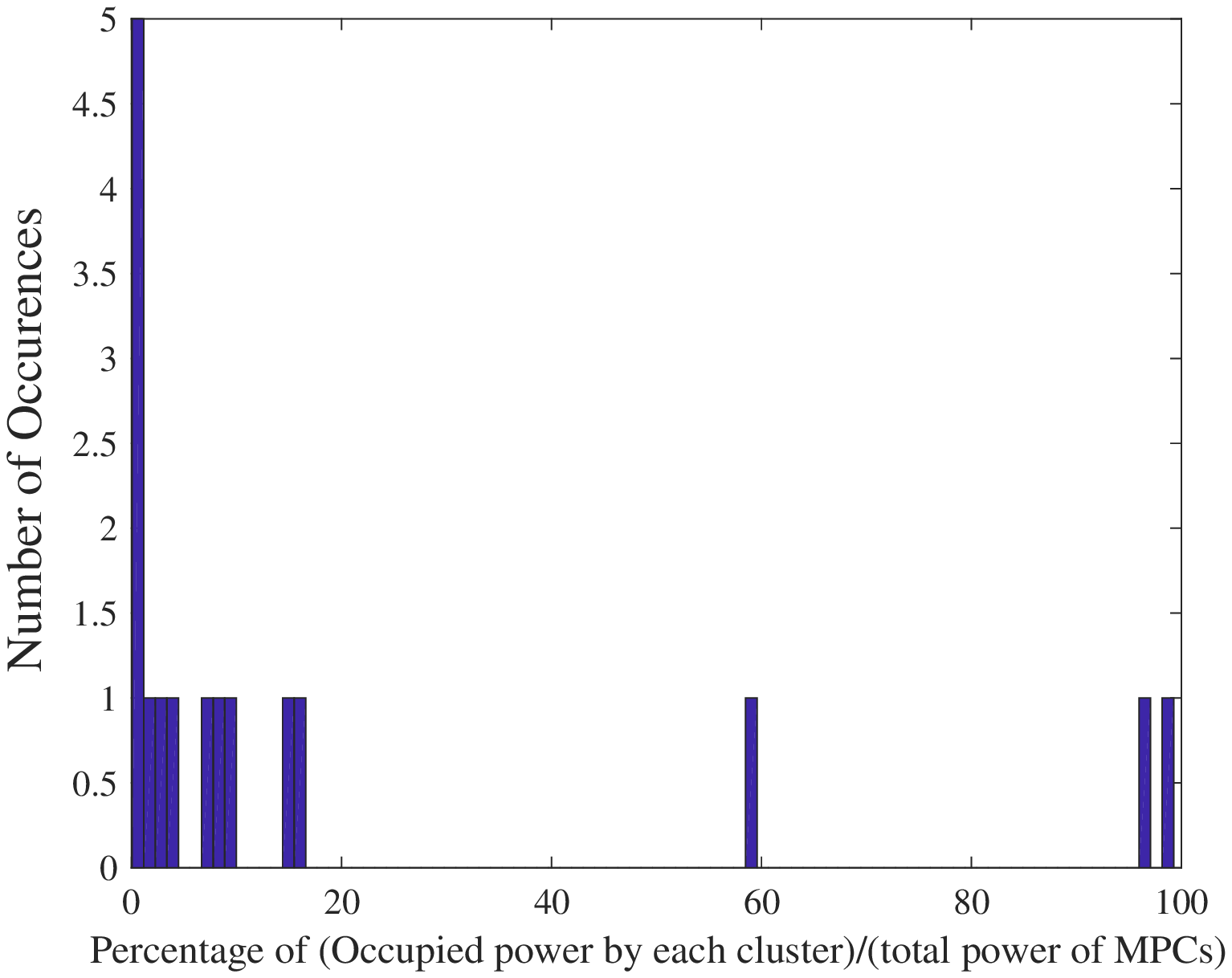}
\vspace{-.2cm}
\caption{Histogram of percentage of occupied power by each Tx-side cluster.}
\label{hist_p_tx}
\end{figure}
\begin{figure}[t!]
\center
\includegraphics[width=82mm]{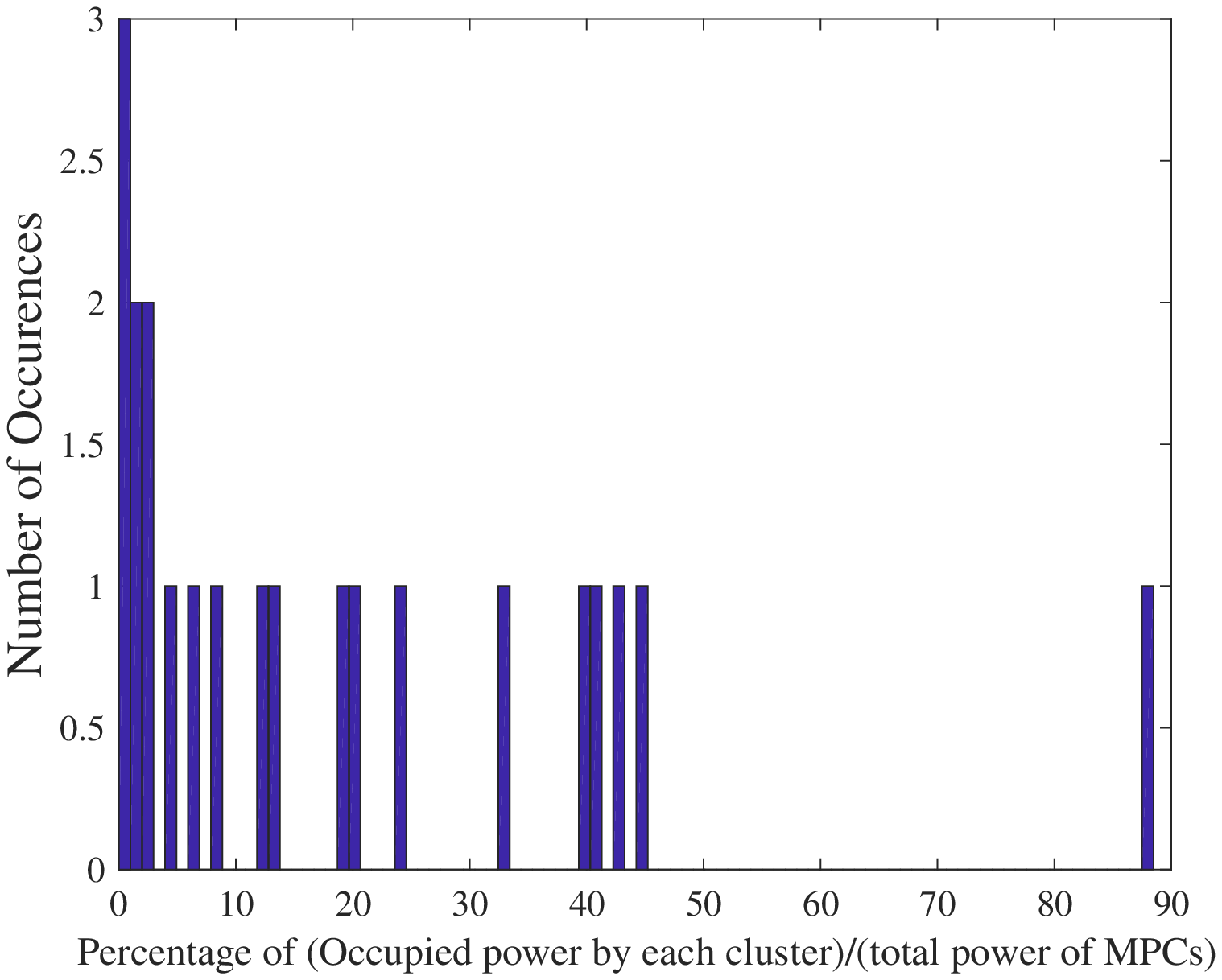}
\vspace{-.2cm}
\caption{Histogram of percentage of occupied power by each Rx-side cluster.}
\label{hist_p_rx}
\end{figure}
The joint clustering-and-tracking algorithm is applied to the ray-tracer results at Helsinki airport, explained in Section \ref{raytracer}, where we have 2639 links. Figs. \ref{cluster_rx_snap1}- and \ref{cluster_tx_snap2} present the exemplary plots for different snapshots. The MPCs are shown by dots, where their power is shown by light blue (weak power) and violet (strong power). The clusters are shown by ellipsoids and always $99\%$ of the total power is carried by the MPCs within clusters. We use different colors for ellipsoids just to make the cluster recognition easier. Each cluster is identified by a cluster ID which is written on each cluster. As these exemplary figures show for snapshots 2,3 and 4, cluster 2 is always tracked while the other clusters are determined as new clusters. 

Next, the lifetime of clusters for the available sets of ray-tracer results is investigated, for Tx-side clusters and Rx-side clusters separately. Figs. \ref{hist_liftetime_rx} and \ref{hist_liftetime_tx} show the histograms of cluster lifetimes for Rx-side (BS-side) and Tx-side (MS-side) scenarios, respectively. The figures show that in most cases clusters are active only for a few snapshots for this set of ray-tracer results. This requires more investigation. Moreover, the number of clusters per snapshot is presented in Figs. \ref{hist_nc_rx} and \ref{hist_nc_tx} for Rx-side and Tx-side clusters, respectively.

The other interesting phenomenon is the movement of the tracked cluster centroids, which is shown in Fig. \ref{center_rx}. Based on these figures the cluster centroids moves rapidly in the $x$ or $y$ direction while its speed is very low in other direction. Moreover, the figure show for these clusters that the centroid's speed is very low in the $z$ direction.
Finally, Figs. \ref{hist_p_tx} and \ref{hist_p_rx} investigate the distribution of the percentage of power in Tx-side and Rx-side clusters.

\section{Conclusions}
In this paper, we have worked on parameterization for the COST 2100 channel model at 60 GHz band. We have worked on a ray-tracer, which has been optimized to match measurements, to get double-directional channels at mmWaves. We have combined clustering and tracking to improve the performance of consistent clustering. The results showed that the joint clustering-and-tracking allows for cluster identification and tracking for the ray-tracer results. Cluster lifetime and number of clusters per snapshot have been investigated. 
\bibliographystyle{IEEEtran}
\bibliography{globe_accept} 

\begin{thebibliography}{10}
\providecommand{\url}[1]{#1}
\csname url@samestyle\endcsname
\providecommand{\newblock}{\relax}
\providecommand{\bibinfo}[2]{#2}
\providecommand{\BIBentrySTDinterwordspacing}{\spaceskip=0pt\relax}
\providecommand{\BIBentryALTinterwordstretchfactor}{4}
\providecommand{\BIBentryALTinterwordspacing}{\spaceskip=\fontdimen2\font plus
\BIBentryALTinterwordstretchfactor\fontdimen3\font minus
  \fontdimen4\font\relax}
\providecommand{\BIBforeignlanguage}[2]{{%
\expandafter\ifx\csname l@#1\endcsname\relax
\typeout{** WARNING: IEEEtran.bst: No hyphenation pattern has been}%
\typeout{** loaded for the language `#1'. Using the pattern for}%
\typeout{** the default language instead.}%
\else
\language=\csname l@#1\endcsname
\fi
#2}}
\providecommand{\BIBdecl}{\relax}
\BIBdecl

\bibitem{multidebbah}
E.~Björnson, E.~A. Jorswieck, M.~Debbah, and B.~Ottersten, ``Multiobjective
  signal processing optimization: the way to balance conflicting metrics in
  {5G} systems,'' \emph{IEEE Signal Process. Mag.}, vol.~31, no.~6, pp. 14--23,
  Oct. 2014.

\bibitem{CiareJointspatial}
A.~Adhikary, E.~A. Safadi, M.~Samimi, R.~Wang, G.~Caire, T.~S. Rappaport, and
  A.~F. Molisch, ``Joint spatial division and multiplexing for {mm-wave}
  channels,'' \emph{IEEE J. Sel. Areas Commun.}, vol.~32, no.~6, pp.
  1239--1255, Jun. 2014.

\bibitem{Molish_tufvesson}
A.~F. Molisch and F.~Tufvesson, ``Propagation channel models for
  {next-generation} wireless communications systems,'' \emph{IEEE Trans.
  Commun.}, vol. {E97-B}, no.~10, pp. 2022--2034, Oct. 2014.

\bibitem{3GPP_spec}
LTE, \emph{Study on channel model for frequency spectrum above 6 {GHz}}.\hskip
  1em plus 0.5em minus 0.4em\relax {3GPP} Specification {38.900}, May 2016.

\bibitem{katsumimocost}
L.~Liu, J.~Poutanen, F.~Quitin, K.~Haneda, F.~Tufvesson, P.~D. Doncker,
  P.~Vainikainen, and C.~Oestges, ``The {COST} 2100 {MIMO} channel model,''
  \emph{IEEE Wireless Commun.}, vol.~19, no.~6, pp. 92--99, Dec. 2012.

\bibitem{ouriet_mic}
M.~Bashar, A.~Burr, K.~Haneda, and K.~Cumanan, ``Robust user scheduling with
  {COST} 2100 channel model for {Massive} {MIMO} networks,'' \emph{IET
  Microwaves, Antennas and Propagation}, vol.~12, no.~11, pp. 1751--8725, Aug.
  2018.

\bibitem{our_ew}
M.~Bashar, A.~G. Burr, D.~Maryopi, K.~Haneda, and K.~Cumanan, ``Robust
  {geometry-based} user scheduling for large {MIMO} systems under realistic
  channel conditions,'' in \emph{Proc. IEEE EW}, May 2018, pp. 1--6.

\bibitem{ourtvt18}
M.~Bashar, A.~Burr, K.~Haneda, and K.~Cumanan, ``Evaluation of low complexity
  {Massive MIMO} techniques under realistic channel conditions,'' \emph{IEEE
  Trans. Veh. Technol.}, Submitted.

\bibitem{Rappaport_globe15}
M.~K. Samimi and T.~S. Rappaport, ``Statistical channel model with multi
  frequency and arbitrary antenna beamwidth for {millimetre-wave} outdoor
  communications,'' in \emph{Proc. {IEEE} {Globecom}}, Dec. 2015.

\bibitem{katsu_vtc15_winner}
A.~Karttunen, J.~Järveläinen, A.~Khatun, and K.~Haneda, ``Radio propagation
  measurements and {WINNER II} parametrization for a shopping mall at {61-65}
  {GHz},'' in \emph{Proc. {IEEE} VTC}, May 2015, pp. 1--6.

\bibitem{katsu_icc16}
K.~Haneda., L.~Tian, H.~Asplund, J.~Li, Y.~Wnag, D.~Steer, C.~Li, T.~Balercia,
  S.~Lee, Y.-S. Kim, A.~Ghosh, T.~Tomas, T.~Nakamura, Y.~Kakishima, T.~Imai,
  H.~Papadopulas, T.~S. Rappaport, G.-R. McCartney, M.~K. Samimi, S.~Sun,
  O.~Koymen, S.~Hur, J.~Park, J.~Zhang, E.~Mellios, A.~F. Molisch, S.~S.
  Ghassamzadeh, and A.~Ghosh, ``Indoor {5G} {3GPP-like} channel models for
  office and shopping mall environments,'' in \emph{Proc. {IEEE} ICC Workshop},
  May 2016, pp. 694--699.

\bibitem{katsu_ref_Iqbal}
C.~Schneider, J.~Gedschold, M.~Kaske, R.~S. Thoma, and G.~D. Galdo,
  ``Estimation and characterization of multipath clusters in urban scenarios,''
  in \emph{Proc. {IEEE} EuCAP}, Apr. 2018, pp. 1--5.

\bibitem{katsu_ref_Schneider}
N.~Iqbal, D.~Dupleich, C.~Schneider, J.~Luo, R.~Muller, S.~Hafner, G.~D. Galdo,
  and R.~S. Thoma, ``Tmodeling of intra-cluster multipaths for 60 {GHz} fading
  channels,'' in \emph{Proc. {IEEE} EuCAP}, Apr. 2018, pp. 1--5.

\bibitem{Gustafson_katsu}
C.~Gustafson, K.~Haneda, S.~Wyne, and F.~Tufvesson, ``On mm-wave multipath
  clustering and channel modeling,'' \emph{IEEE Trans. Ant. Prop.}, vol.~62,
  no.~3, pp. 1445--1455, Mar. 2014.

\bibitem{katsu_airpot_vtc18}
K.~Haneda, J.~Järveläinen, and A.~Karttunen, ``[online]. available:
  https://arxiv.org/pdf/1802.08591.pdf,'' in \emph{Proc. {IEEE} VTC}, 2018, pp.
  1--6.

\bibitem{katsu_trans_16_laser}
J.~Järveläinen, K.~Haneda, and A.~Karttunen, ``Indoor propagation channel
  simulations at 60 ghz using point cloud data,'' \emph{IEEE Trans. Ant.
  Prop.}, vol.~64, no.~8, pp. 4457--4467, Aug. 2016.

\bibitem{nicki_track_china06}
N.~Czink, R.~Tian, S.~Wyne, F.~Tufvesson, J.~P. Nuutinen, J.~Ylitalo, E.~Bonek,
  and A.~F. Molisch, ``Tracking {time-variant} cluster parameters in {MIMO}
  channel measurements,'' in \emph{Proc. CHINACOM}, Aug. 2007, pp. 1147--1151.

\bibitem{nicki_indoor_china06}
N.~Czink, E.~Bonek, L.~Hentila, J.~P. Nuutinen, and J.~Ylitalo,
  ``{Cluster-based} {MIMO} channel model parameters extracted from indoor
  {time-variant} measurements,'' in \emph{Proc. IEEE Globecom}, Nov. 2006, pp.
  1--5.

\bibitem{kay}
S.~Kay, \emph{Fundamentals of statistical signal processing: Estimation
  theory}.\hskip 1em plus 0.5em minus 0.4em\relax Englewood Cliffs, NJ, USA,
  Prentice-Hall, 1993.

\bibitem{nicki_letter_mcd}
N.~Czink, P.~Cera, J.~Salo, E.~Bonek, J.~P. Nuutinen, and J.~Ylitalo,
  ``Improving clustering performance using multipath component distance,''
  \emph{Electronics Letters}, vol.~42, no.~1, pp. 1--2, Jan. 2006.

\bibitem{nicki_vtc06_clustering_frame}
N.~Czink, P.~Cera, J.~Salo, E.~Bonek, J.~p.~Nuutinen, and J.~Ylitalo, ``A
  framework for automatic clustering of parametric {MIMO} channel data
  including path powers,'' in \emph{Proc. {IEEE} VTC}, Sep. 2006, pp. 1--5.

\end{thebibliography}
\end{document}